\documentclass[runningheads]{llncs}

\usepackage{subcaption}
\usepackage{graphicx}
\usepackage[hidelinks]{hyperref}
\usepackage{xcolor}

\newcommand\blfootnote[1]{%
	\begingroup
	\renewcommand\thefootnote{}\footnote{#1}%
	\addtocounter{footnote}{-1}%
	\endgroup
}

\title{Word-level style control for expressive, non-attentive speech synthesis}
\author{Konstantinos Klapsas \inst{1}\and Nikolaos Ellinas\inst{1}\and  June Sig Sung\inst{2}\and Hyoungmin Park\inst{2}\and Spyros Raptis\inst{1} }

\authorrunning{K. Klapsas et al.}

\institute{ Innoetics, Samsung Electronics, Greece \and  
Mobile Communications Business, Samsung Electronics, Republic of Korea \\
\email{   k.klapsas@partner.samsung.com \\
\{n.ellinas, js6.sung, hm94.park, s.raptis\}@samsung.com}}

\begin{document}

\maketitle
\begin{abstract}
This paper presents an expressive speech synthesis architecture for modeling and controlling the speaking style at a word level. It attempts to learn word-level stylistic and prosodic representations of the speech data, with the aid of two encoders. The first one models style by finding a combination of style tokens for each word given the acoustic features, and the second outputs a word-level sequence conditioned only on the phonetic information in order to disentangle it from the style information. The two encoder outputs are aligned and concatenated with the phoneme encoder outputs and then decoded with a Non-Attentive Tacotron model. An extra prior encoder is used to predict the style tokens autoregressively, in order for the model to be able to run without a reference utterance. We find that the resulting model gives both word-level and global control over the style, as well as prosody transfer capabilities. 
  
\end{abstract}
\noindent\textbf{Keywords}: text-to-speech, prosody control, expressive speech synthesis

\section{Introduction}

\blfootnote{The final authenticated publication is available online at \url{https://doi.org/10.1007/978-3-030-87802-3_31}}

Since neural text-to-speech models such as Tacotron \cite{tacotron,tacotron2} significantly improved the quality and naturalness of speech synthesis systems, there has been an increased interest in approaches that expand on the basic architecture by explicitly modeling the prosody of the synthesized speech. This can be done either by using a reference from which prosody is copied \cite{pmlr-v80-skerry-ryan18a}, or by manually controlling the prosody on an utterance \cite{Wang2018StyleTU}, or fine-grained level \cite{Lee2019RobustAF}.  

Of particular interest to our work is the Global Style Tokens mechanism (GST) \cite{Wang2018StyleTU} which assigns for each utterance a style embedding that is learned in an unsupervised way as a combination of a given number of Style Tokens. A precursor to GST \cite{2017arXiv171100520W}, has a similar architecture, where the tokens are extracted at a phoneme level. Our approach is an extension of this architecture, where the tokens are extracted at the word level, something that provides more refined control than the GST architecture yet it operates at a more intuitive level than the phoneme level.

A new family of models that has recently gained traction is the Non-Attentive family \cite{Shen2020NonAttentiveTR,durian} which, instead of using attention to align the phoneme encoder outputs with the target spectrogram, employs extra modules to directly predict the duration of each phoneme. The ground-truth durations are used during training both as a target for the prediction, and as the input for the decoder. This removes a number of artifacts that are typically observed in attention models, allowing for a more immediate and robust control of duration related properties, such as speaking rate. We leverage this by conditioning the duration prediction module to the style embeddings, thereby gaining the ability to control the speaking rate via the tokens that comprise the embeddings.

\subsection{Related Work}
A number of approaches based on VAEs have been successful in learning latent representations in an unsupervised manner
\cite{hsu2018hierarchical,Zhang2019LearningLR}.  While this is usually done at the utterance level, fine-grained approaches that align the target mel-spectrogram with the phonetic features via an attention mechanism have also been explored. In \cite{fully_hierarchical}, after this alignment is obtained, a hierarchical VAE structure is used where phonetic information is conditioned on word-level aggregated  acoustic features, while in \cite{sun2020generating} a quantization module is used in order to discretize the latent information. 

While VAE methods have the advantage that it is possible to synthesize audio without a reference by manipulating the latent variables directly, systems which require a given reference audio have also been investigated, aiming to copy the prosody from a different speaker \cite{Karlapati2020CopyCatMF,Klimkov2019FinegrainedRP}. The features extracted from the reference audio are aligned at the phoneme level in order to achieve fine-grained prosody transfer.

There have also been several approaches that, similarly to ours, aggregate the information at a word level \cite{pmlr-v97-kenter19a,Hono2020HierarchicalMG}. Most of these works condition the style embeddings purely on the textual information instead of the acoustic features, as is done in the original formulation of GSTs. Additionally, a number of these models condition on linguistic embeddings that are derived from pre-trained text models such as BERT \cite{bert1,bert2}.
\subsection{Our Contributions}
Our main contributions in this paper are:
\begin{itemize}
\item we modify the GST architecture to achieve word-level prosody control conditioned on acoustic features.
\item we propose a word sequence encoder that helps disentangle the text content from style information.
\item we incorporate a prior autoregressive encoder \cite{sun2020generating} that allows synthesis without the need for a reference, without compromising the control capabilities of the model.
\item we investigate and demonstrate the control abilities of such configuration in a non-attentive architecture.

\end{itemize}

\section{Method}

\begin{figure*}[t]
  \centering
  \includegraphics[width=\textwidth]{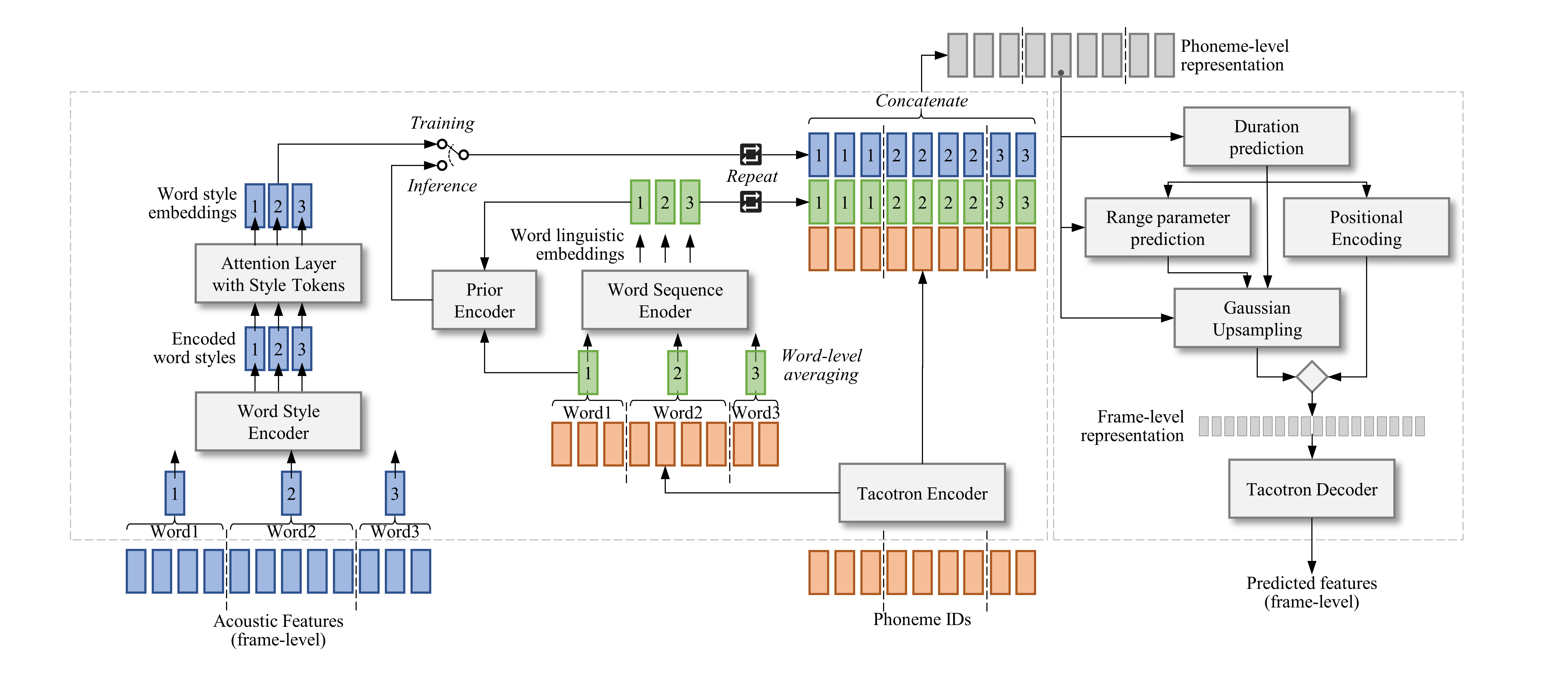}
  \caption{Architecture of full model.}
  \label{fig:full model}
\end{figure*}

\subsection{Word Style Encoder}

The frame-level acoustic features are first aligned with the phonemes of the training utterances using a forced-alignment system \cite{2016ExpressiveSS}. We employed an HMM monophone acoustic model trained using flat start initialization and implemented with the HTK toolkit \cite{htk}, similarly to ASR forced alignment models. After the alignments are obtained for each utterance in the training set, the duration of each phoneme is extracted and then used to train the prediction module of the Non-Attentive Tacotron. The phonemes are then aligned with the words, using the known phone-word alignment. 

For each word, an encoder summarizes the corresponding acoustic sequence into a fixed length vector that contains the prosodic information for the word. 
This vector is then used as a query to an attention module and a weighted average of a fixed number of style tokens is obtained via softmax. Similar to \cite{Wang2018StyleTU}, the tokens are initialized randomly and their values are updated by back propagation. 

The output of the softmax is the word style embedding that is then replicated for each phoneme of the word, concatenated with the Non-Attentive Tacotron encoder outputs and sent to the decoder for reconstruction. By concatenating the style tokens with the phoneme encoder outputs, we essentially condition the duration prediction module on the style. We therefore expect the style information to contain information about the speaking rate, and indeed we find that some tokens correspond directly to the speaking rate. We elaborate more on the interpretability of the tokens in Section \ref{interpretability}.

This model requires a reference utterance, since it learned during training to expect a valid combination of different tokens for each word, and generates unintelligible speech if a single token is used. However, it is possible to increase the weight the model assigns to any token manually, thus biasing the style of any particular word towards the given token. This can be done by adding the embedding vector of the given token to the full embedding matrix, as is done in \cite{2017arXiv171100520W}. This process can be done to all the words of the sentence, thereby gaining global style control, or it can be done to any individual word, using the default embeddings from the ground truth for the rest of the words. 

To increase  the  robustness of the process of directly controlling the token weights, we estimate the distribution of each token's weights in the training corpus and then apply changes to the token weights that are multiples of their standard deviation. We  find  that  this offers a  unified way  to  control weights  across  tokens, inducing style changes that are perceptually similar in intensity for all tokens.

\subsection{Word Sequence Encoder}

For the model architecture as described, it was observed that the word style embeddings contain a lot of phonetic information. In order to disentangle the style and phonetic representation of each word, a second word-level encoder was used, conditioned only on phonetic information. 

This encoder takes as input the Tacotron encoder outputs and, after they are passed through a linear layer, their average is extracted for every word. 
The obtained sequence, is passed through a bidirectional word-level LSTM and concatenated with the encoder outputs and the word style embeddings. Like the style tokens and style encoder, it is only trained through the reconstruction loss. A stop gradient operation is used so that the Tacotron encoder does not get updated by the gradient that passes through the word sequence encoder.

The output of the module contains information that is already present on the text but aggregates it at the same level as the word style encoder. Therefore, the decoder learns to rely on the style embeddings only for style information since the remaining phonetic information is already modeled.

\subsection{Prior autoregressive encoder} \label{Prior autoregressive encoder}
In order to synthesize without a reference utterance, a prior autoregressive encoder similar to the one proposed in \cite{sun2020generating} is used to generate a sequence of word style embeddings, given only the textual information. This model is trained to predict each word style embedding given the text and all the previous embeddings.

The prior encoder takes as inputs the outputs of both the phoneme encoder and the word sequence encoder.The phoneme encoder output is first averaged across each word and concatenated with the output of the word sequence encoder. A stop gradient operator is used for both other encoders, so that only the parameters of the prior model are updated. An extra rnn cell is used which, given this word-level sequence, predicts the word embeddings autoregressively by maximizing the likelihood of the embedding sequence.
%
%


%

%

The embeddings are predicted in the same space as the output of the token attention module, ie this model tries to predict the weighted combination of the tokens at once. This is similar to the TPSE-GST architecture from \cite{Stanton2018PredictingES} where they predict the global style tokens from textual information. During inference the encoder requires only the phoneme encodings and a vector of zeros that is fed as the initial state of the rnn cell.

The parameters of the Tacotron model and the extra word-level encoders are not updated with the loss of this prior encoder. Therefore, it can be trained after the rest of the model is done training or simultaneously, without affecting the rest of the training procedure. It was observed, however, that training the models simultaneously leads to better performance.

During inference, the prior encoder predicts plausible style embeddings which are fairly rich  in expressiveness. Moreover, it is possible to directly manipulate the prior embeddings by adding weighted style token vectors at any word's embedding, thus biasing those words towards a specific desired style.


We find that using the prior encoder in place of the reference does not significantly deteriorate nor the quality nor the expressiveness of the produced speech. Furthermore, no additional fine-tuning of the token weights is required, since the weights that correspond to meaningful changes in style are the same ones as those with a given reference.

The architecture of the full model, including all three encoders can be seen in Fig. \ref{fig:full model}.

\section{Experiments}
\subsection{Experimental Setup}
The models were evaluated on a subset of the Blizzard Challenge 2013 single-speaker audiobook dataset  \cite{King2014TheBC} which contains 85 hours of speech with highly varying prosody.  All audio data was first resampled to 24 kHz before the extraction of acoustic features. 

Similar to \cite{LPCtron}, the acoustic features that were used for these experiments are 22 LPC features, i.e.
20 Bark-scale cepstral coefficients, the pitch period and the pitch correlation, that are then vocoded using the LPCNet Vocoder \cite{lpcnet}. We use a custom implementation of the Non-Attentive Tacotron decoder model \cite{Shen2020NonAttentiveTR} in which the durations of each phoneme are predicted given the encoder outputs and then a Gaussian upsampling is applied in order to get the correct length of the acoustic features. A content-based attention \cite{luong-etal-2015-effective} with a softmax is used to predict the weights of 15 style tokens. The number 15 was chosen empirically as the maximum number of tokens that can be used without the decoder learning to ignore a large number of them. The size of the token embeddings is 128, also found via experimentation.

We use the Adam optimizer \cite{Kingma2015AdamAM}  with parameters (0.9, 0.999) for training the networks with batch size 32. We use the same learning rate schedule as the Non-attentive Tacotron, with a linear ramp-up for 4K steps and then decay at half every 50K steps. We also apply L2 regularization with factor $10^{-6}$. 

All of the following evaluations were done on a subset of 1000 utterances from the dataset that were not seen by the model during training.

Audio samples and illustrations from the experiments are available at \url{https://innoetics.github.io/publications/word-style-tokens/index.html}. 

\subsection{Reconstruction Performance} \label{recon}
We use $F_0$ frame error (FFE), Voice Decision Error (VDE) and Gross Pitch Error (GPE) \cite{FFE} as well as mel-cepstral distortion (MCD) \cite{MCD} as objective evaluation metrics for the reconstruction performance. We compare our model with the baseline Non-Attentive Tacotron both when we reconstruct from a given sentence and when we synthesize using the embeddings predicted by the prior encoder. DTW was used to align the ground-truth sequences with the outputs of each model. The results are shown in Table \ref{tab:mcd_ffe}.

\begin{table}[th]
  \caption{Objective evaluation scores for reconstruction. Lower is better.}
  \label{tab:mcd_ffe}
  \centering
  \begin{tabular}{l l l l l}
  \hline
    \textbf{Model} & \multicolumn{1}{c}{\textbf{FFE}} & \multicolumn{1}{c}{\textbf{VDE}} & \multicolumn{1}{c}{\textbf{GPE}} & \multicolumn{1}{c}{\textbf{MCD}}\\
  \hline
    Non Attentive Tacotron            & $30.3$ & $6.1$ & $34.1$ & $5.8$ ~~~             \\
    Proposed model with prior                   & $34.5$  & $6.5$ & $39.0$ & $5.9$ ~~~            \\
    Proposed model with reference               & $11.5$  & $4.9$ & $9.3$ & $4.9$ ~~~            \\
\hline
  \end{tabular}
\end{table}

As expected the reconstruction is significantly better for the model when using the reference embeddings. The performance using the prior embeddings is only slightly worse than that of the baseline Non Attentive Tacotron, indicating that the prior does indeed learn how to capture an average style of a sentence.

\subsection{Subjective Evaluation} \label{subjective}
We evaluate the naturalness of our samples with a Mean Opinion Score (MOS), ranging from 1 to 5. We compare samples from the baseline Non-Attentive Tacotron with the samples synthesized with the prior embeddings, and with the reference embeddings. We also include scores for samples biased towards four tokens from the prior embeddings, for three different biasing weights, that correspond to 1, 2 and 4 standard deviations from the mean of the respective tokens, in either direction. The tokens were chosen for their clear semantic interpretation (further explored in section \ref{interpretability})  and the weights were chosen to range from minor but perceptible changes to very clear changes while keeping the speech intelligible. The samples were crowdsourced and raters who assigned unusually low ratings to ground truth samples were excluded from the analysis, leaving 38 raters.

The results are shown in Table \ref{tab:mos}. The quality of the speech synthesized from the prior model is only slightly worse than the baseline Non-Attentive Tacotron, which could  be partly attributed  to  the  fact  that  the latter seems to model prosody more conservatively. Furthermore, while the quality degrades as larger biases are imposed towards specific tokens, as is expected, the generated speech is still perceived as mostly natural, even when the changes are substantial.
   
\begin{table}[th]
  \caption{MOS Evaluation scores}
  \label{tab:mos}
  \centering
  \begin{tabular}{l l }
  \hline
    \textbf{Model} & \multicolumn{1}{c}{\textbf{MOS}}   \\
 \hline
    Natural speech (ground truth)            & $4.41$         \\
    Non Attentive Tacotron            & $4.33$             \\
    Proposed with prior                 & $4.27$   ~~~            \\
    Proposed with reference                 & $4.28$   ~~~            \\
    Proposed with token weight $-4$ stds          & $3.82$   ~~~            \\
    Proposed with token weight  $-2$ stds         & $4.26$   ~~~            \\
    Proposed with token weight  $-1$ stds        & $4.27$   ~~~            \\
    Proposed with token weight  $+1$ stds         & $4.30$   ~~~            \\
    Proposed with token weight  $+2$ stds         & $4.16$   ~~~            \\
    Proposed with token weight  $+4$ stds         & $4.03$   ~~~            \\
   \hline
  \end{tabular}
\end{table}

\subsection{Objective assessment of model's generative behavior 
}
The generative behavior of the model has also been examined in terms of the acoustic properties of its generated audio files. Three models have been examined: 
\begin{enumerate}
\item a plain Non-Attentive Tacotron model without any extra component for modeling styles, 
\item a variation of our proposed model where the input of the Prior Encoder is simply the word-level aggregated phoneme encoder outputs,
\item our proposed model as seen in Fig. \ref{fig:full model} where the Prior Encoder gets as additional input the outputs of the Word Sequence Encoder, as described in \ref{Prior autoregressive encoder}
\end{enumerate}

The properties examined were the pitch and phoneme durations. Inference was performed using each of the three models and the synthesized audio was segmented using an acoustic model trained on the full dataset. From there, phoneme durations were obtained which were then z-normalized per phone-class using statistics from the full dataset.

\begin{figure}
\centering
\begin{subfigure}{.33\textwidth}
  \centering
  \includegraphics[width=\linewidth]{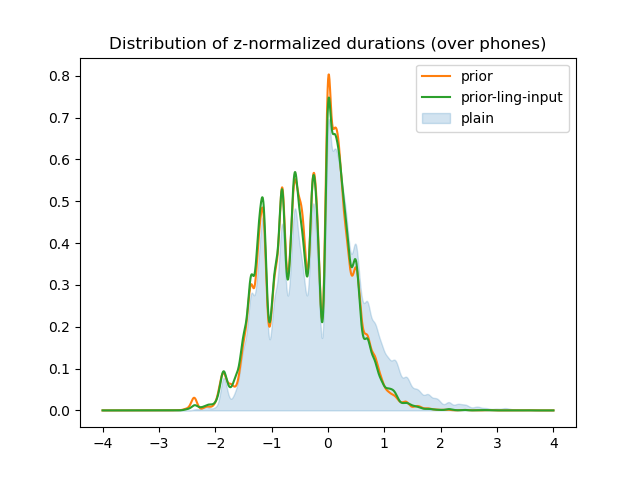}
  \caption{Durations}
  \label{fig:obj1}
\end{subfigure}%
\begin{subfigure}{.33\textwidth}
  \centering
  \includegraphics[width=\linewidth]{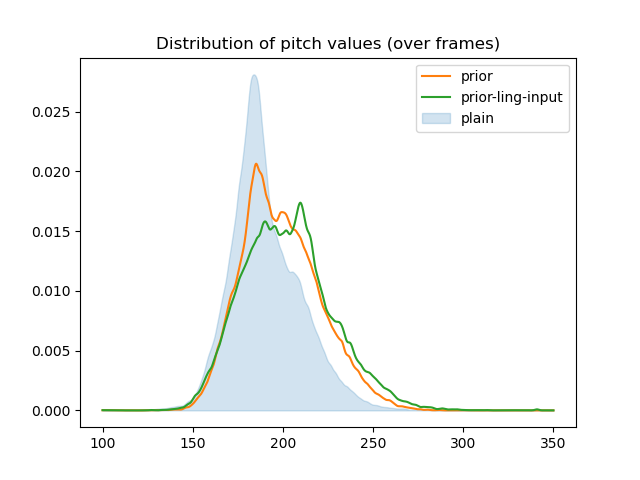}
  \caption{Pitch values}
  \label{fig:obj2}
\end{subfigure}
\begin{subfigure}{.33\textwidth}
  \centering
  \includegraphics[width=\linewidth]{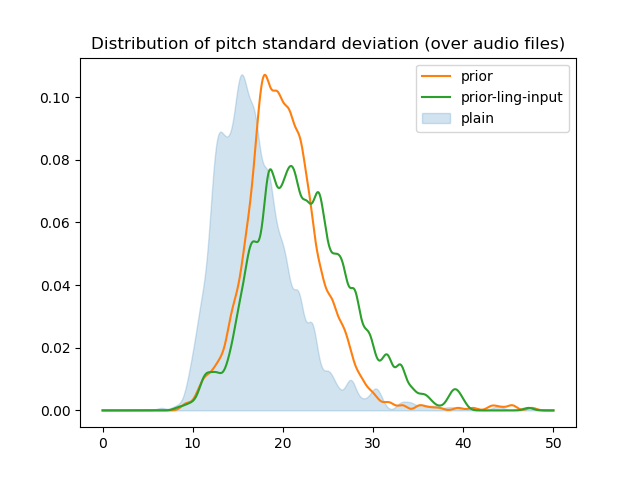}
  \caption{Pitch deviations}
  \label{fig:obj3}
\end{subfigure}%
\caption{Distributions of relevant characteristics}
\label{fig:obj}
\end{figure}

The distribution of these normalized duration values for all phonemes in each of the three generated datasets is shown in Fig. \ref{fig:obj1}. The pitch contours of all generated files were calculated on a frame basis (excluding unvoiced frames). These were used to calculate the distributions of: (i) these raw values (Fig. \ref{fig:obj2}); as well as (ii) the standard deviation of pitch per audio file (Fig. \ref{fig:obj3}). Kernel-density estimation was used to estimate each of the distributions above, using Gaussian kernels and empirically selected bandwidths in each case. 

As evident from these figures, the phoneme durations (Fig. \ref{fig:obj1}) in the audio generated by Models 2 and 3 which include style-related components are similar to those of the plain Model 1. The style-aware Models 2 and 3 tend to generate speech with pitch spanning a higher range (Fig. \ref{fig:obj2}), Model 3 even more so compared to Model 2. They also tend to generate higher pitch variation within each generated utterance (Fig. \ref{fig:obj3}). 

These indicate that the style-aware models tend to generate richer pitch patterns than the plain model. This behavior may be partly attributed to the larger size of these models and their, thus, increased modeling capacity. 

\subsection{Interpretability of tokens} \label{interpretability}
A general weakness of unsupervised methods such as GSTs and VAEs is that the representations of style that the models learn depends heavily on training factors such as initialization,  hyperparameters etc. However, we find that across all our experiments, a number of the tokens had simple intuitive interpretations. In particular, we find that some tokens were directly related to the pitch and some to the speaking rate and that this fact was generally consistent and independent of the details of each particular training. Additionally, we often find that by adding the embedding of a particular token with a negative weight corresponds to the conceptually opposite change than by adding the embedding with a positive weight. For example, the token that corresponds to higher speaking rate, has a slow-down effect when subtracted rather than added to the style embedding. This is consistent with the findings in \cite{Wang2018StyleTU}.

\begin{figure}
\centering
\begin{subfigure}{.5\textwidth}
  \centering
  \includegraphics[width=\linewidth]{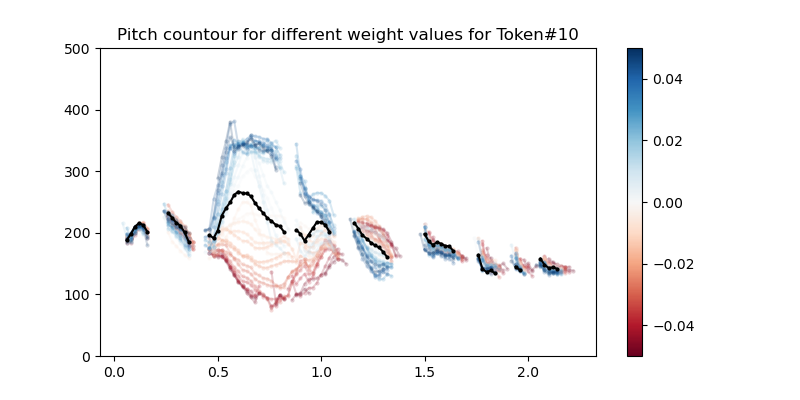}
  \caption{One Word}
  \label{fig:sub1}
\end{subfigure}%
\begin{subfigure}{.5\textwidth}
  \centering
  \includegraphics[width=\linewidth]{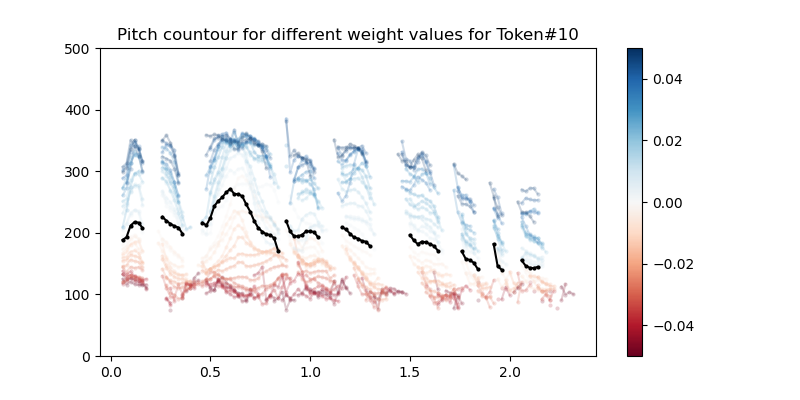}
  \caption{All words}
  \label{fig:sub2}
\end{subfigure}
\caption{Increasing/decreasing a pitch-controlling token}
\label{fig:interp1}
\end{figure}

\begin{figure}
\centering
\begin{subfigure}{.5\textwidth}
  \centering
  \includegraphics[width=\linewidth]{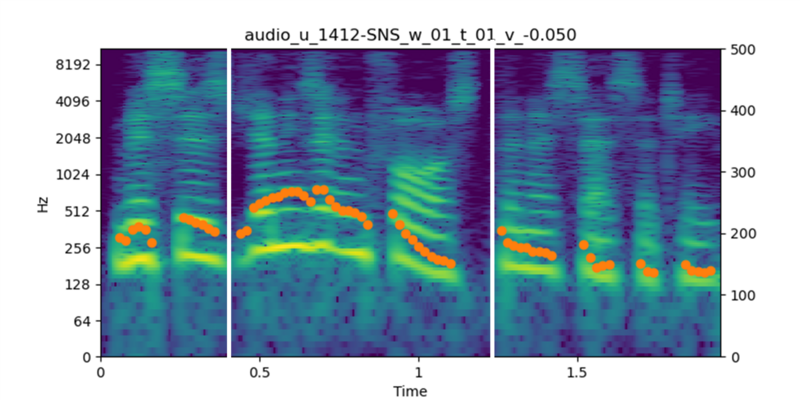}
  \caption{Negative weight}
  \label{fig:sub1}
\end{subfigure}%
\begin{subfigure}{.5\textwidth}
  \centering
  \includegraphics[width=\linewidth]{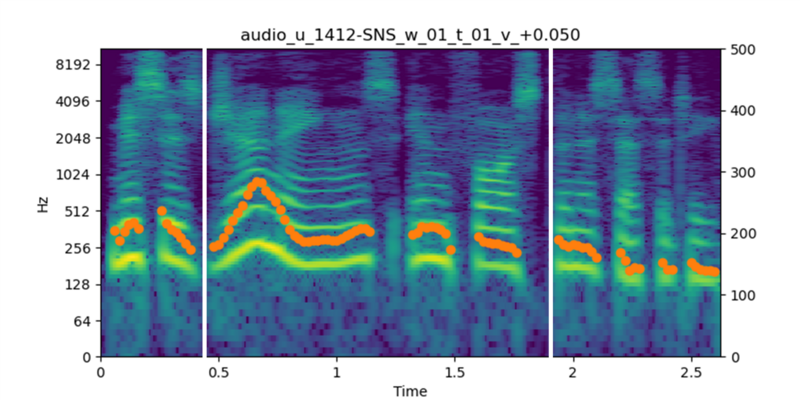}
  \caption{Positive weight}
  \label{fig:sub2}
\end{subfigure}
\caption{Increasing/decreasing a duration-controlling token}
\label{fig:interp2}
\end{figure}

In Fig \ref{fig:interp1}, we see the difference of a token that controls the pitch when added with a positive or a negative sign to the embedding matrix. In Fig. \ref{fig:interp2} we see the same for a duration controlling token.

\subsection{Style transfer}
We also find that it is possible to do style transfer from a different given sentence (or word) by using the embeddings from that target sentence. The transfer is more successful when the number of phonemes in the source and target words are similar. 

We observe that despite the fact that the two encoders are disentangling to some extend the phonetic information from the style information, the naive implementation of style transfer still sometimes results to unintelligible words. We address this by synthesizing using a mixture of the weights of the prior encoder and the style of the target sentence. Since the prior encoder predicts the style embeddings given the phonemes of the target sentence, it results to more intelligible speech, by reducing somewhat the transfer ability.

\begin{figure}[h]
\centering
\begin{subfigure}{.5\textwidth}
  \centering
  \includegraphics[width=\linewidth]{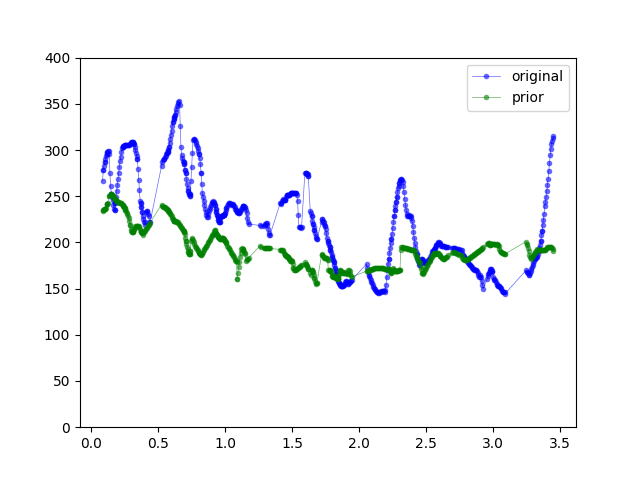}
  \caption{From prior embeddings}
  \label{fig:sub1}
\end{subfigure}%
\begin{subfigure}{.5\textwidth}
  \centering
  \includegraphics[width=\linewidth]{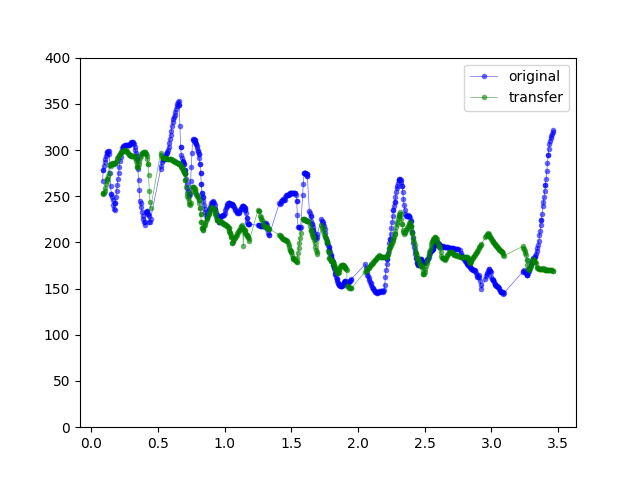}
  \caption{From given embeddings}
  \label{fig:sub2}
\end{subfigure}
\caption{Style Transfer}
\label{fig:transfer}
\end{figure}

Figure \ref{fig:transfer} shows the $F_0$ plots of the target sentence compared with the sentence synthesized from prior embeddings, and with the embeddings of the given sentence. We find that indeed the style transfer does manage to match the desired $F_0$.

\section{Conclusions}
As demonstrated in the experiments, the proposed model offers word-level style modeling with fairly robust control capabilities. Many of the emergent style tokens seem well correlated to perceptually relevant and interpretable aspects of the speaking style, also making it possible to transfer styles across words and utterances in a meaningful way. This is further enhanced by the proposed model's non-attentive nature which virtually removes any repeat/omit failures that are typical in Tacotron-like architectures. Some keys directions for further research will focus on integrating pre-trained linguistic word embeddings and investigating alternative ways to meaningfully control the style.  

\newpage

\bibliographystyle{splncs04}
\bibliography{mybib}

\end{document}